\documentclass[11pt,twocolumn,aps,prb,reprint]{revtex4-1}

\usepackage{graphicx}
\usepackage{amsmath}
\begin{document}
\title{Hydrodynamic model for electron-hole plasma in graphene}

\author{D. Svintsov$^1$, V. Vyurkov$^1$,  S. Yurchenko$^2$, T. Otsuji$^{3,5}$, and
V. Ryzhii$^{4,5}$}

\affiliation
{$^1$ Institute of Physics and Technology, Russian Academy of Sciences,
Moscow 117218, Russia} 
\affiliation
{$^2$ Center for Photonics and Infrared Engineering,
Bauman Moscow State University,  
Moscow 105005, Russia} 
\affiliation
{$^3$ Research Institute for Electrical Communication, 
Tohoku University, Sendai 980-8577, 
 Japan} 
\affiliation
{$^4$ Computational Nanoelectronics Laboratory, University of Aizu,
Aizu-Wakamatsu 965-8580, Japan}
\affiliation
{$^5$ Japan Science and Technology Agency, CREST, Tokyo 107-0075, Japan}

\begin{abstract}
We propose a hydrodynamic model describing steady-state and dynamic electron and hole transport properties of graphene structures which accounts for the features of the electron and hole spectra. It is intended for electron-hole plasma in graphene characterized by high rate of inter-carrier scattering compared to external scattering (on phonons and impurities), i.e., for intrinsic or optically pumped (bipolar plasma), and gated graphene (virtually monopolar plasma). We demonstrate that the effect of strong interaction of electrons and holes on their transport can be treated as a viscous friction between the electron and hole components. 
We apply the developed model for the calculations of the graphene dc conductivity, in particular, the effect of mutual drag of electrons and holes is described. 
The spectra and damping of collective excitations in graphene in the bipolar and monopolar limits  are found. It is shown that at high gate voltages and, hence, at high electron and low hole densities (or vice-versa), the excitations are associated with the self-consistent electric field and the hydrodynamic pressure (plasma waves). In intrinsic and optically pumped graphene, the  waves constitute quasineutral perturbations of the electron and hole densities (electron-hole sound waves) with the velocity being dependent only on the fundamental graphene constants. 

\end{abstract}
\maketitle

\section{Introduction}
The hydrodynamic approach is quite reasonable for the description of dense electron-hole plasma in semiconductor systems in which the electron-electron, electron-hole, and hole-hole collisions dominate over the collisions of electrons and holes with disorder. 
In particular, such a situation can occur in intrinsic graphene at the room temperature, particularly, in the structures with low or moderate permittivity of the layers between which the graphene layer is clad, when the characteristic inter-carrier collision frequency can reach high values. As a result,  this frequency can be  much greater than that of collisions with impurities or phonons. Similar situation occurs in the gated graphene at sufficiently high gate voltages when large electron or hole densities can be induced as well as at strong optical pumping of graphene.
The hydrodynamic models  of the electron-hole systems in different  structures and devices based on the standard semiconductors with parabolic and near-parabolic energy spectra of electrons and holes are widely used (see, for instance, Ref.~\cite{Jungel}). However, in case of graphene, such models and the pertinent equations should be revised due to the linear energy spectra of electrons and holes. A hydrodynamic approach was recently used to describe the stationary transport processes in graphene~\cite{MacDonald}. However, the role of a strong electron-hole scattering, which, as shown in our work, is crucial, was not addressed.

In this paper, we develop a strict hydrodynamic model for electron-hole plasma in graphene and demonstrate its workability in some applications. The paper is organized as follows. In Sec.~II, we derive the hydrodynamic equations (continuity equations, Euler equations, and energy transfer equations) for graphene from the Boltzmann-Vlasov kinetic equations for massless electrons and holes assuming high carrier-carrier collision frequencies.  Section~III deals with the application of the derived equations for the calculations of graphene dc conductivity taking into account the effect of electron-hole drag. In Sec.~IV, the collective excitations in graphene at different conditions are considered using the obtained hydrodynamic equations. In particular, we show that two types of weakly damping excitations can exist in the electron-hole plasma: (1) electron (or hole) plasma waves in gated graphene in the state with  the  Fermi level far from   the Dirac point and (2) quasi-neutral electron-hole sound waves in the bipolar electron-hole plasma. The damping of both plasma waves (in virtually monopolar plasma) and electron-hole sound waves (in bipolar plasma) is determined by the scattering on disorder, while the electron-hole scattering almost does not affect the damping. In contrast, the plasma waves in bipolar electron-hole systems exhibit a strong damping due to electron-hole scattering processes. In Sec.~V, we draw the main conclusions. Some cumbersome formulas are singled out to the Appendix.  

\section{Derivation of hydrodynamic equations}
For the massless electrons in holes in graphene the spectrum is linear $\varepsilon(p)=v_Fp$. Hence, the kinetic equations governing the distribution functions $f_e = f_e({\bf p})$ and $f_h = f_h({\bf p})$ read, respectively,
\begin{multline}
\label{Kin_el}
\frac{\partial f_e}{\partial t}+v_F\frac{{\bf p}}{p}\frac{\partial f_e}{\partial{\bf r}} + e \frac{\partial \varphi}{\partial \bf r} \frac{\partial f}{\partial {\bf p}}=\\
St\{f_e,f_e\}+St\{f_e,f_h\} + St_{i}\{f_e\},
\end{multline}
\begin{multline}
\label{Kin_hole}
\frac{\partial f_h}{\partial t}+v_F\frac{{\bf p}}{p}\frac{\partial f_h}{\partial{\bf r}}- e \frac{\partial \varphi} {\partial \bf r}\frac{\partial f}
{\partial {\bf p}}= \\
St\{f_h,f_h\}+St\{f_h,f_e\}+St_{i}\{f_h\}.
\end{multline}
Here $v_F{\bf p}/p=\partial \varepsilon (p)/\partial{\bf p}$ is the electron (hole) velocity, $v_F \simeq 10^8$ cm/s is 
the characteristic velocity of electrons and holes in graphene (Fermi velocity),  $e = |e|$ is the absolute value of the electron charge, ${\bf E}=-\partial\varphi/\partial{\bf r}$ is the electric field, $St_i\{f_e\}$ and $St_i\{f_h\}$
are the collision integrals of electrons and holes, respectively, with  disorder (impurities and phonons);  $St\{f_e,f_e\}$, $St\{f_h,f_h\}$, and  $St\{f_e, f_h\}$ are the inter-carrier collision integrals. In Eqs.~(\ref{Kin_el}) and (\ref{Kin_hole}) we have neglected the recombination terms assuming  that the recombination rate is much smaller than that of collisions and the frequency of the plasma waves under consideration.

As known~\cite{Gantmacher}, the Fermi distribution functions of electrons and holes
\begin{equation}
\label{F_el}
f_{e}({\bf p})=\left[1+\exp\left( \frac{\varepsilon(p)-{\bf p}\cdot{\bf V}_e-\mu_e}{T}
\right)\right]^{-1},
\end{equation}
\begin{equation}
\label{F_hole}
f_{h}({\bf p})=\left[1+\exp\left( \frac{\varepsilon(p)-{\bf p}\cdot{\bf V}_h+\mu_h}{T}
\right)\right]^{-1}
\end{equation}
turn the electron-electron and hole-hole collision integrals to zero  owing to the conservation of momentum and energy in the inter-carrier collisions. In Eqs.~(\ref{F_el}) and~(\ref{F_hole}) ${\bf V}_e$ and ${\bf V}_h$ are the average (drift) velocities of electrons and holes, respectively, $\mu_e$ and $\mu_h$ are electron and hole chemical potentials, and the temperature $T$ is measured in energy units.

Hereafter we consider rather small drift velocities and perform an expansion of Eqs.~(\ref{F_el}) and~(\ref{F_hole}) over~${\bf{V}}_e$ and~${\bf{V}}_h$:
\begin{equation}
\label{F_lin_el}
f_{e}({\bf p})= f_{e,0}-\frac{\partial f_{e,0}}{\partial \varepsilon}{\bf p}\cdot{\bf V}_e,
\end{equation}
\begin{equation}
\label{F_lin_hole}
f_{e}({\bf p})= f_{h,0}-\frac{\partial f_{h,0}}{\partial \varepsilon}{\bf p}\cdot{\bf V}_h.
\end{equation}
Here $f_{e,0}$ and $f_{h,0}$ stand for the functions, given by Eqs.~(\ref{F_el}) and~(\ref{F_hole}), with 
${\bf V}_e = {\bf V}_h = 0$. The distribution functions (\ref{F_lin_el},\ref{F_lin_hole}) still turn the electron-electron and hole-hole collision integrals to zero and, at the same moment, conserve the number of particles (the  density is the same in moving and stationary frames).

If the electric field is sufficiently weak and the inhomogenity of the electron-hole plasma is small (as it will be assumed in the following), the electron-hole collision integrals can be expanded over ${\bf V}_e$ and ${\bf V}_h$ and presented as 
$St\{f_e,f_h\}=- St\{f_h,f_e\}\simeq ({\bf V}_h - {\bf V}_e)\cdot {\bf S}\{f_{e,0}, f_{h,0} \}$ (for details see the Appendix).
In this form, the electron-hole collision terms describe the friction between the electron and hole plasma components. Similarly, the terms corresponding to the collisions of electrons and holes with disorder can be presented in the form
$St_i\{f_e\}  = -  {\bf V}_e \cdot {\bf S}_i\{f_{e,0}\}$
and 
$St_i\{f_h\}  = - {\bf V}_h \cdot {\bf S}_i\{f_{h,0}\}$.
Here ${\bf{S}}$ and ${\bf S}_i$ are the functionals of the distribution functions $f_{e,0}$ and $f_{h,0}$.

At small values of electron and hole average velocities ${\bf V}_e$ and ${\bf V}_h$, the friction terms can be considered as perturbations in comparison with the electron-electron and hole-hole collision terms. Thus the distribution functions given by Eqs.~(\ref{F_el}) and~(\ref{F_hole}) are the approximate solutions of Eqs.~(\ref{Kin_el}) and (\ref{Kin_hole}). Then the quantities ${\bf V}_e $, ${\bf V}_h$, $\mu_e$, and $\mu_h$ (or the electron and hole sheet densities, $\Sigma_e$ and $\Sigma_h$) can be found considering the terms in the left-hand sides of Eqs.~(\ref{Kin_el}) and (\ref{Kin_hole}) and the friction terms as perturbations, using the standard procedure (akin to the Chapman-Enskog method~\cite{Ford-Uhlenbeck} for the derivation of the hydrodynamic equations from the kinetic equations). 

On integrating Eqs.~(\ref{Kin_el}) and~(\ref{Kin_hole}) over $d\Gamma_{\bf p}=g d^2 {\bf p}/(2 \pi \hbar)^2$ (where $g=4$ is the electron degeneracy factor in graphene), we obtain the continuity equations for electrons and holes
\begin{equation}
\label{Cont}
\frac{\partial \Sigma_e}{\partial t}
+\frac{\partial \Sigma_e {\bf V}_e}{\partial\bf r}=0 ,
\qquad
\frac{\partial \Sigma_h}{\partial t}
+\frac{\partial \Sigma_h {\bf V}_h}{\partial\bf r}=0.
\end{equation}

To derive the Euler equations, one should integrate Eqs. (\ref{Kin_el}) and (\ref{Kin_hole}) times $\bf p$ over $d\Gamma_{\bf p}$. In the case of parabolic dispersion, one could multiply the Boltzmann equation either by velocity or momentum as they are proportional to each other. For linear dispersion the choice of momentum is crucial. Just the momentum is conserved in particle-particle collisions unlike to the velocity which can be changed. After integration one obtains the system of Euler equations for electrons and holes:
\begin{multline}
\label{Euler_el}
\frac{3}{2}
\frac{\partial}{\partial t}
\frac{\left\langle {p_e}\right\rangle {{\bf{V}}_{e}}}{v_F}
+\frac{\partial }{\partial \bf {r}}\frac{v_F \left\langle {p_e} 
\right\rangle }{2}- e {\Sigma_e}\frac{\partial \varphi}
{\partial \bf{r}}=\\
-\beta_e {{\bf V}_e}-{\beta_{eh}}\left( {\bf V}_e-{\bf V}_h \right), 
\end{multline}
\begin{multline}
\label{Euler_hole}
\frac{3}{2}
\frac{\partial }{\partial t}
\frac{\left\langle {p_h} \right\rangle {{\bf{V}}_h}}{v_F}+\frac{\partial }{\partial \bf{r}}\frac{{v_F}\left\langle {p_h} \right\rangle }{2}+ e {\Sigma_{h}}\frac{\partial \varphi }{\partial \bf {r}}=\\
-{\beta_h}{{\bf V}_h}-{\beta_{eh}}\left( {{\bf V}_h}-{{\bf V}_e} \right).
\end{multline}
Here the angle brackets denote an integration over the equilibrium Fermi distribution functions, in particular,
$$
\langle p_e \rangle =\int_0^{\infty} \left[1+\exp\left(\displaystyle\frac{v_F p-\mu_{e}}{T}\right)\right]^{-1}\frac{2\pi g p^2 dp}{(2\pi\hbar)^2}
$$
is the momentum modulus per unit area, the friction coefficients $\beta_{eh}$, $\beta_e$ and $\beta_h$ are the functions of the non-perturbed (steady-state) values of the chemical potentials $\mu_{e,0}$ and $\mu_{h,0}$. 

One can rewrite Euler equations in a classical form on introducing the fictitious carrier masses
$$
M_{e}=\frac{\langle p_{e} \rangle}{v_F \Sigma_{e}},\qquad
M_{h}=\frac{\langle p_{h} \rangle}{v_F \Sigma_{h}},
$$ 
which are estimated as $0.016$ of the free electron mass at $\mu_e= \mu_h = 0$ and $T=300$ K.

The detailed derivation of the friction coefficients can be found in the Appendix; here we write down only the final expressions in several limits. The electron-hole friction coefficient $\beta_{eh}$ can be represented as
\begin{equation}
\label{Beta_eh}
\beta_{eh} = A \frac{T^4e^4}{\hbar^5v_F^6\kappa^2} \cdot I\left( \frac{\mu_e}{T}, \frac{\mu_h}{T}\right).
\end{equation}
where $\kappa$ is the effective permittivity of environment (the substrate and gate dielectric), the dimensionless constant $A$ of the order of unity and the function $I\left( \mu_e/T, \mu_h/T\right)$ can be obtained after proper linearization of the electron-hole collision integral (see the Appendix). For intrinsic graphene $I\left( 0, 0\right)=1$, while for monopolar plasma in gated graphene  $I\left( \mu/T, \mu/T\right) \propto e^{-\mu/T}$ due to the exponentially small number of holes. In the current paper, we shall adopt the following interpolation for $\beta_{eh}$ which is valid in the limiting cases of monopolar plasma and intrinsic graphene: 
\begin{equation}
\label{Beta_eh_interp}
\beta_{eh}=\frac{\nu_{eh}}{v_F}\frac{\langle p_e\rangle \langle p_h\rangle}{\langle p_e\rangle + \langle p_h\rangle}=\nu_{eh}\frac{\Sigma_eM_e\Sigma_hM_h}{\Sigma_eM_e+\Sigma_hM_h}.
\end{equation}
Here we have introduced the electron-hole collision frequency $\nu_{eh}$ to be estimated below in the Sec.~III.

When the acoustic phonon scattering dominates, the coefficients $\beta_e$ and $\beta_h$ can be presented as~\cite{Vasko-Ryzhii, Phonon-limited-conductivity}
$$
\beta_e=\frac{D^2T\langle p_e^2\rangle}{4\rho s^2 \hbar^3 v_F^2},
\qquad\beta_h=\frac{D^2T\langle p_h^2\rangle}{4\rho s^2 \hbar^3 v_F^2},
$$
where $D$ is the deformation potential constant, $\rho$ is the sheet density of graphene, $s$ is the sound velocity. Due to the existence of several acoustic phonon branches and considerable discrepancy in experimental data of graphene constants~\cite{Vasko-Ryzhii, Falkovsky-Sound, Phonon-limited-conductivity} it is reasonable to use semi-phenomenological formulas:
\begin{equation}
\label{Phonon_scat}
\beta_e=\lambda T\langle p_e^2\rangle,\qquad 
\beta_h=\lambda T\langle p_h^2\rangle,
\end{equation}
and extract the numerical value of $\lambda$ from experimental data on dc conductivity~(see Sec.~III). 

For scattering on charged impurities, the friction coefficients $\beta_e$ and $\beta_h$ are reasonably proportional to the carrier densities and the density of charged impurities $\Sigma_i$
\begin{equation}
\label{Imp_scat}
\beta_e \propto  \Sigma_e \Sigma_i,\qquad
\beta_h \propto  \Sigma_h \Sigma_i.
\end{equation}
In general, the density of charged impurities depends on the chemical potentials and temperature.

At last, the system of hydrodynamic equations should be supplemented with the energy transfer  equations. Introducing the energy density $\langle \varepsilon_{e,h}\rangle=v_F\langle p_{e,h}\rangle$, the pertinent equations can be written down as
\begin{equation}
\label{Energy_el}
\frac{\partial \langle \varepsilon_e\rangle}{\partial t}+\frac{3}{2}\frac{\partial \langle \varepsilon_e \rangle {\bf V}_e}{\partial {\bf r}}-e \Sigma_e {\bf V}_e\frac{\partial \varphi}{\partial {\bf r}}=Q_e,
\end{equation}
\begin{equation}
\label{Energy_hole}
\frac{\partial \langle \varepsilon_h\rangle}{\partial t}+\frac{3}{2}\frac{\partial \langle \varepsilon_h \rangle {\bf V}_h}{\partial {\bf r}}+ e \Sigma_h {\bf V}_h\frac{\partial \varphi}{\partial {\bf r}}=Q_h,
\end{equation}
where the heat sink rates $Q_e$ and $Q_h$ are proportional to the difference between electron and phonon temperatures. However, in the present paper we neglect the heating as we deal with low-field dc conductivity and high-frequency plasma waves.

\section{Effect of electron-hole drag and dc conductivity of gated graphene}
As one of the demonstrations of the hydrodynamic equations applications we calculate the dc conductivity of gated graphene. It is assumed that the electric potential and the charge density are related by the Poisson equation. In the gradual channel approximation~\cite{Shur}
\begin{equation}
\label{Gradual_ch}
C(V_G  - \varphi) = e(\Sigma_e  - \Sigma_h).
\end{equation}
Here  $C=\kappa/4\pi d$ is the specific capacitance per unit area, $d$ and $\kappa$ are the thickness and permittivity of the gate dielectric, respectively, and $V_G$ is the gate voltage.

To calculate the dc conductivity we rewrite the Euler equations for the steady-state situation in terms of electrochemical potentials:
\begin{equation}
\label{Steady_el_mu}
{\Sigma_e} \frac{\partial \left( {\mu_e}- e \varphi  \right)}{\partial \bf r}=-{\beta_e}{{\bf V}_{e}}-{\beta_{eh}}\left( {{\bf V}_e}-{{\bf V}_h} \right),
\end{equation}
\begin{equation}
\label{Steady_hole_mu}
{\Sigma_h}\frac{\partial \left( e \varphi -{\mu_h} \right)}{\partial \bf r}=-{\beta_h}{{\bf V}_h}-{\beta_{eh}}\left( {{\bf V}_h}-{{\bf V}_e} \right),
\end{equation}
where the derivative of electric potential is associated with drift current, as well as the derivative of chemical potential is associated with diffusion current. In the following we restrict our calculations by the consideration of  the drift current only and, therefore, omit the terms $\partial\mu_{e,h}/\partial{\bf r}$ in Eqs.~(\ref{Steady_el_mu}) and (\ref{Steady_hole_mu}).  As a result, we arrive at the following expressions for the mean (drift) velocities of electrons and holes:
\begin{equation} 
\label{Drift_vel}
{\bf V}_e = -\biggl[\frac{(\Sigma_e-\Sigma_h)\beta_{eh}+\beta_h\Sigma_e}
{\beta_{eh} (\beta_e+\beta_h)+\beta_e \beta_h}\biggr]\, 
\frac{\partial \varphi}{\partial \bf r},
\end{equation}
\begin{equation}
{\bf V}_h = \biggl[\frac{(\Sigma_h-\Sigma_e)\beta_{eh}+\beta_e\Sigma_h}{\beta_{eh} (\beta_e+\beta_h)+\beta_e \beta_h}\biggr]\,
 \frac{\partial \varphi}{\partial \bf r}.
\end{equation}
Substituting the quantities ${\bf V}_e$ and ${\bf V}_h$ from Eqs.~(\ref{Drift_vel},21) into the general expression for current density 
${\bf j} = e (\Sigma_h {\bf V}_h-\Sigma_e {\bf V}_e)= 
G\,(-\partial \varphi/\partial {\bf r})$,
one obtains the following formula for the conductivity~$G$:
\begin{equation}
\label{Conductivity_gen}
G= \frac{e^2 {{\left( \Sigma_e-\Sigma_h \right)}^2}}{{\beta_e}+{\beta_h}+\beta_e\beta_h / \beta_{eh}}+\frac{e^2\left( \Sigma_e^2 \beta_h+ \Sigma_h^2 \beta_e\right)}{{\beta_{eh}}\left( \beta_e+\beta_h \right)+ \beta_e\beta_h} .
\end{equation}

The first term in the right-hand side of Eq.~(\ref{Conductivity_gen}) is associated with the scattering of electrons and holes on impurities and phonons. This term turns into zero at the Dirac point. However, far from the Dirac point, i.e., in purely electron or hole plasma, this term dominates. Meanwhile, the second term is due to the contribution of the  electron-hole friction. The latter results in a high resistivity of graphene at the Dirac point and its vicinity. 

To calculate the graphene minimal conductivity, we reasonably assume that at the Dirac point scattering between electrons and holes prevails: $\beta_{e}\ll\beta_{eh}$, $\beta_{h}\ll\beta_{eh}$. Plugging Eq.~(\ref{Beta_eh}) for $\beta_{eh}$ into Eq.~(\ref{Conductivity_gen}) one obtains an expression for intrinsic graphene conductivity
\begin{equation}
\label{Conduct_DP2}
G_0=\frac{e^2 \Sigma_0^2}{\beta_{eh}} \propto \frac{\hbar v_F^2 \kappa^2}{e^2}.
\end{equation}
Strikingly, the intrinsic graphene conductivity (minimum conductivity) does not depend on temperature. This is in agreement with the experimental results~\cite{Novoselov} demonstrating a constant conductivity over a broad range of temperature from 0.3~K to 300~K in which the carrier density varies by 6 orders of magnitude. Worth mentioning the Eq.~(\ref{Conduct_DP2}) was previously obtained using the  scaling theory~\cite{Vyurkov-Ryzhii}, and also via thorough description of electron-hole scattering~\cite{Kashuba}. 

Earlier, the minimum conductivity of graphene was calculated under the assumption of strong interaction among carriers~\cite{Zitterbewegung, Ziegler-Kubo, Falkovsky-Varlamov}. The strength of interaction is governed by the "fine-structure constant" $\alpha = e^2/\kappa\hbar v_F$ which is equal to 2.2 for intrinsic (suspended) graphene ($\kappa=1$). The strong-interaction theories bind the conductivity of graphene to the conductance quantum $e^2/h$ regardless of the permittivity of environment. As the parameter $\alpha$ is about unity, both theories of weak and strong interaction give rise to close values of conductivity. However, in prospective graphene structures the screening caused by dielectrics and nearby gates make the value of $\alpha$ less than unity, therefore, the weak-interaction theories become more adequate to the situation.

After substituting Eq.~(\ref{Beta_eh_interp}) for $\beta_{eh}$ into Eq.~(\ref{Conduct_DP2})one can rewrite that equation in the conventional form appropriate for the estimation of the electron-hole collision frequency $\nu_{eh}$:
\begin{equation}
\label{Conduct_DP1}
G_0=\frac{2\Sigma_0 e^2}{M_0 \nu_{eh}},
\end{equation}
where $M_0$ is the fictitious mass of electrons and holes in intrinsic graphene dependent on temperature. Comparing $G_0$ given by Eq.~(\ref{Conduct_DP1}) with the experimental value $G_0 \simeq$ (6 kOhm)$^{-1}$, one can estimate the electron-hole collision frequency as $\nu_{eh}=3 \times 10^{13}$ s$^{-1}$. Such a high value justifies an employment of the hydrodynamic model.

In the monopolar limit ($\Sigma_e\gg\Sigma_h$), the expression for the conductivity could be also simplified:
\begin{equation}
\label{Conduct_MP}
G=\frac{e^2 \Sigma_e^2}{\beta_e}.
\end{equation}
Equation~(\ref{Conduct_MP}) allows to estimate the coefficient $\lambda$ in the phonon collision term given by Eq.~(\ref{Phonon_scat}). It is widely assumed that in suspended graphene samples or twisted graphene stacks the conductivity and mobility are limited by phonon scattering only. For instance~\cite{Bolotin_T-dependent}, the carrier mobility $B$, defined as
$$
B=\frac{G}{e\left|\Sigma_e-\Sigma_h\right|},
$$
reaches the value of 120,000~cm$^2~$V$^{-1}$~s$^{-1}$ at $\Sigma_e-\Sigma_h=2\times 10^{11}$ cm$^{-2}$ and the temperature $T=240$ K (when the chemical potential $\mu \simeq 0.028$ eV). Hence, in the monopolar limit, one obtains
\begin{equation}
\label{Mobility}
B\simeq \frac{e}{\beta_e \Sigma_e} = \frac{e \Sigma_e}{\lambda T \langle p_e^2\rangle}.
\end{equation}
Equation~(\ref{Mobility}) yields $\lambda\simeq 2.4 \times 10^{54}$ cm~s$^{-1}~$J$^{-2}$. Accordingly,  the characteristic collision frequency $\nu_{e}=2v_F\beta_e/3\langle p_e\rangle$ at room temperature varies from $8.6\times 10^{11}$~s$^{-1}$ at the Dirac point to $3.5\times 10^{12}$~s$^{-1}$ at $V_G=10$~V and $d=10$~nm. 

\begin{figure}
\includegraphics[width=1\linewidth]{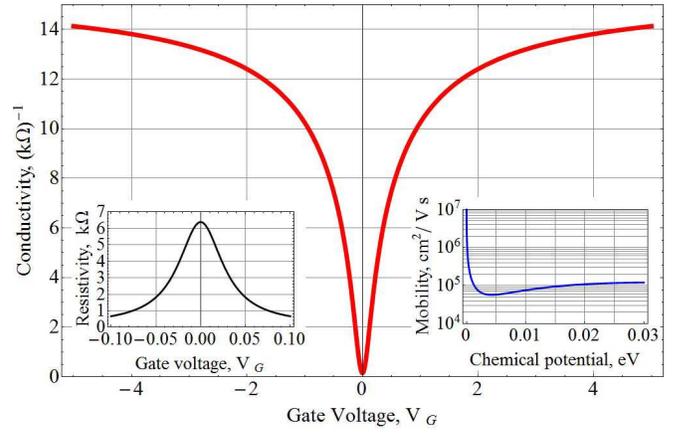}
\caption{\label{Fig1}Conductivity of graphene $G$ vs. gate voltage at $T=300$ K, $d=10$ nm, and $\kappa=4$. Insets: left panel - resistivity $G^{-1}$ near the Dirac point vs. gate voltage, right panel - mobility $B$ vs. chemical potential.}
\end{figure}

Figure~\ref{Fig1} demonstrates the dependence of the low-field dc conductivity $G$ of graphene on the gate voltage $V_G$, plotted using Eqs.~(\ref{Conductivity_gen}, \ref{Beta_eh_interp}, \ref{Phonon_scat}). 
The voltage dependence of the graphene resistivity and the dependence of the carrier mobility on the chemical potential $\mu$ [given by Eq.~(\ref{Mobility})] are shown in the insets in Fig.~\ref{Fig1}. In the case when scattering on acoustic phonons dominates, the conductivity tends to saturation at high gate voltages and the phonon-limited mobility decreases. To describe the conductivity of realistic graphene structures, one should account for other scattering objects: charged impurities, interfacial phonons, bulk phonons, etc. For example, scattering on charged impurities results in the following trends of conductivity curves in monopolar case: $G(V_G) \propto V_G$ if the density of impurities is constant and $G(V_G) \propto \sqrt{V_G}$ if the distribution of impurity levels is uniform in an energy scale~\cite{Vyurkov-Ryzhii}. The latter fact manifests a decrease in defect scattering at low temperatures.

The conductivity and mobility curves in Fig. 1 exhibit a good agreement with the experimental works, where phonon-limited mobility was investigated~\cite{Bolotin_Mobility,Bolotin_T-dependent,Shishur}. As for the theoretical approaches to the description of graphene conductivity, significant efforts have been focused on the calculation of the minimum conductivity and the conductivity far from the neutrality point~\cite{Vasko-Ryzhii,Phonon-limited-conductivity, Intrinsic-Mobility}. The conductivity at an arbitrary value of carrier density was considered in Refs.~\cite{Falkovsky-Varlamov, Ziegler-robust-transport} without specifying the mechanism of scattering. Our model provides an opportunity to calculate the transport characteristics of graphene at any charge carrier density taking into account the definite scattering processes. Above all, we show that the effect of electron-hole drag significantly impacts the transport properties of graphene and leads to an abrupt drop of graphene resistivity outside of the Dirac point.

It is remarkable that the drag effect is crucial even at a small mismatch in the electron and hole densities \cite{Vyurkov-Ryzhii}. The conductivity in the regime of drag almost does not depend on electron-hole scattering, although it still remains the strongest one in the system. Indeed, the phonon (impurity) scattering term in the expression for conductivity~(\ref{Conductivity_gen}) dominates under the condition
\begin{equation}
\label{Drag}
\frac{|\Sigma_h-\Sigma_e|}{\Sigma_e} > \sqrt{\frac{\beta_{h}} {\beta_{eh}} \biggr|}_{\mu=0}.
\end{equation}
For the preceding estimations, the ratio given by Eq.~(\ref{Drag}) is evaluated as 0.3 at room temperature. Providing ineq.~(\ref{Drag}) is satisfied, the difference between electron and hole velocities (Eq.~\ref{Drift_vel}) becomes rather small:
\begin{equation}
\frac{|V_h-V_e|}{V_h+V_e}\simeq \frac{1}{2} \sqrt{\frac{\beta_{h}} {\beta_{eh}} \biggr|}_{\mu=0}.
\end{equation}
The difference between velocities still exists because of the opposite charges of electrons and holes and, therefore, the opposite directions of the electric forces. In principle, for the drag regime the two-fluid description of graphene system might be reduced to a single electron-hole fluid with renormalized charge of particles according to Eq.~(\ref{Drag}). However, it seems preferable to retain two-fluid model to cover all possible situations.

It is readily seen from above mentioned that a narrow resistivity peak in Fig. 1 could be explained by the drag effect. Moreover, stronger scattering leads to a wider resistivity peak. The latter is in concordance with the experimental data~\cite{Bolotin_Mobility}.

\section{Plasma and electron-hole sound waves in graphene}
\subsection{General dispersion relation for collective excitations}
Below we demonstrate the application of the hydrodynamic equations for the calculation of the spectra of collective excitations in electron-hole system. We use the same gradual channel approximation as in the previous section [Eq.~(\ref{Gradual_ch})].
To derive the spectra, one can apply the common technique of small-signal analysis assuming that
\begin{gather}
\mu_{e}=\mu_{e,0}+\delta \mu_{e}e^{{i(kx-\omega t)}},\qquad
\mu_{h}=\mu_{h,0}+\delta \mu_{h}e^{{i(kx-\omega t)}},\nonumber \\
\delta V_{e}=\delta V_{e,0}e^{{i(kx-\omega t)}},\qquad
\delta V_{h}=\delta V_{h,0}e^{{i(kx-\omega t)}},\nonumber \\
\varphi=\varphi_0+\delta \varphi e^{{i(kx-\omega t)}}, \nonumber
\end{gather}
where $\delta \mu_{e}$, $\delta \mu_{h}$, $\delta V_{e,0}$, $\delta V_{h,0}$, and $\delta \varphi$ are the amplitudes of alternating variations. The non-perturbed chemical potentials $\mu_{e,0}$ and $\mu_{h,0}$ coincide if the steady state is an equilibrium one and are  determined by the doping and the gate potential. In particular, at $V_G = 0$ and zero doping the Fermi level is located in the Dirac point. In intrinsic graphene under optical interband pumping the chemical potentials of electrons and holes can be rather different. For example, at $V_G = 0$, $\mu_{e,0} = \mu_0 > 0$ and $\mu_h = -\mu_0 < 0$, where $\mu_0$ is determined by the intensity of pumping.

From the linearized versions of Eqs.~(\ref{Cont}-\ref{Euler_hole}) as well as Eq.~(\ref{Gradual_ch}) the following system of algebraic equations arises:
\begin{widetext}
\begin{equation}
\label{Wave_dynamics_el}
\left[ -i\omega \frac{3}{2}\frac{\left\langle {p_e} \right\rangle }{v_F}+\frac{ik^2 \Sigma_e^2}{\omega}\left( \frac{v_F}{\left\langle p_e^{-1} \right\rangle }+\frac{e^2}{C} \right)+{\beta_e}+{\beta_{eh}} \right]\delta {{V}_{e}}-\left[ \frac{i{k^2}{\Sigma_e} {\Sigma_h}}{\omega }\frac{e^2}{C}+{\beta_{eh}} \right]\delta {V_h}=0, 
\end{equation}
\begin{equation}
\label{Wave_dynamics_hole}
\left[ -i\omega \frac{3}{2}\frac{\left\langle p_h \right\rangle }{v_F}+\frac{i k^2 \Sigma_h^2}{\omega }\left( \frac{v_F}{\left\langle p_h^{-1} \right\rangle}+\frac{e^2}{C} \right)+{\beta_h}+{{\beta }_{eh}} \right]\delta {V_h}-\left[ \frac{i{k^2}{\Sigma_e} {\Sigma_h}}{\omega }\frac{e^2}{C}+{\beta_{eh}} \right]\delta {V_e}=0.
\end{equation}

The solvability condition for Eqs.~(\ref{Wave_dynamics_el}) and (\ref{Wave_dynamics_hole}) results in the general dispersion relation for the collective excitations:
\begin{multline}
\label{General_dispersion}
\left[ -i\omega +\frac{ik^2 v_e^2}{\omega}\left( 1+ r_e \right)+\nu_{e}+{v_F}\frac{2\beta_{eh}}{3\left\langle p_e \right\rangle} \right]\left[ -i\omega +\frac{ik^2 v_h^2}{\omega}\left( 1+ r_h \right)+\nu_{h}+{v_F}\frac{2\beta_{eh}}{3\left\langle p_h \right\rangle } \right]=\\
=\left[ \frac{i k^2 v_e^2 \Sigma_h}{\omega \Sigma_e}r_e+\frac{\beta _{eh}v_F}{\langle p_e\rangle} \right] \left[ \frac{i k^2 v_h^2 \Sigma_e}{\omega \Sigma_h}r_h+\frac{\beta _{eh}v_F}{\langle p_h\rangle} \right].
\end{multline}
\end{widetext}
Here for brevity we have introduced 
the dimensionless constants 
$$
r_e=\displaystyle\frac{e^2 \left\langle p_e^{-1} \right\rangle }{C v_F}, \qquad 
r_h=\displaystyle\frac{e^2\left\langle p_h^{-1} \right\rangle}{C v_F},
$$
the squared characteristic velocities 
$$
v_e^2=\displaystyle\frac{2\Sigma_e^2 v_F^2}
{3\left\langle {p_e}\right\rangle \left\langle p_e^{-1} \right\rangle}, \qquad 
v_h^2=\displaystyle\frac{2\Sigma_h^2 v_F^2}{3\left\langle p_h \right\rangle \left\langle p_h^{-1} \right\rangle},
$$
and the frequencies 
$$
\nu_{e}=\displaystyle\frac{2v_F\beta_e}{3\langle p_e\rangle}, \qquad
\nu_{h}=\displaystyle\frac{2v_F\beta_h}{3\langle p_h\rangle}.
$$
The quantities $r_e$ and $r_h$ determine the ratio of the electrostatic energy to the kinetic energy in the wave. In the particular gated structures we have considered above ($\kappa=4$, $d=10$~nm) the dimensionless constant $r_e$ varies from 1.9 at $V_G=0$ to 45 at $V_G=10$~V. The electron-phonon collision frequencies $\nu_e$ and $\nu_h$, as shown below, determine the damping of the waves.

\subsection{Analytical solutions for symmetric bipolar and monopolar systems}
Exact solutions of the general dispersion equation~(\ref{General_dispersion}) can be acquired for symmetric bipolar plasma (optically pumped or intrinsic graphene) and monopolar plasma.

In symmetric bipolar plasma all quantities characterizing the electron and hole systems coincide (e.g. $v_e=v_h=v$, $\Sigma_e=\Sigma_-=\Sigma$, ...), therefore, we shall omit the subscripts in this certain case. In symmetric systems Eq.~(\ref{General_dispersion}) provides two solutions:

\begin{equation}
\label{Sound_branch}
\omega_{-} =-i\frac{\nu}{2}+\sqrt{{{k}^{2}}v^{2}- \left(\frac{\nu}{2}\right)^2}.
\end{equation}
\begin{equation}
\label{Plasma_branch_DP}
\omega_{+} =-i\left(\frac{\nu}{2}+\frac{\nu_{eh}}{3}\right)+\sqrt{{{k}^{2}}v^{2}(1+2r)^2- \left(\frac{\nu}{2}+\frac{\nu_{eh}}{3}\right)^2}.
\end{equation}

The second solution $\omega_+$ represents the waves with opposite motion of electrons and holes leading to the strong damping. These waves correspond to the perturbations of charge density (plasma waves).

The branch given by Eq.~({\ref{Sound_branch}}) is of the most interest. Electrons and holes in these waves move in the same direction, i.e., the electron-hole plasma remains quasi-neutral. Due to the co-directional motion of the electron and hole components, their mutual collisions do not affect the wave damping. In the case, it originates from the collisions of electrons and holes with impurities and phonons. We can conclude that those waves are associated solely with the pressure gradient. In the following, such waves will be referred to as electron-hole sound waves in analogy with electron-ion sound waves in classical plasma.

The velocity of electron-hole sound is
\begin{equation}
\label{Sound_velocity}
s_{-} = v = v_F \sqrt{\frac{2\Sigma^2}{3\left\langle {p}\right\rangle \left\langle p^{-1}\right\rangle}}.
\end{equation}
It does not depend on structure parameters and does not exceed the Fermi velocity for any electron and hole densities.  If the Fermi level crosses the Dirac point the analytical expression for velocity is
$$
s_-=v_F\frac{\pi^2}{18 \sqrt{\ln 2\zeta(3)}} \simeq 0.6 v_F,
$$
where $\zeta(x)$ stands for Riemann zeta function.

The dispersion laws for monopolar plasma can be derived providing the inequalities $\Sigma_e \gg\Sigma_h$, $\beta_e\gg \beta_h$, and $r_e\gg 1\gg r_h$ are satisfied:
\begin{equation}
\label{Plasma_branch}
{\omega_{+}}=-i \frac{\nu_{e}}{2} + \sqrt{k^2v_e^2(1+r_e)-\left( \frac{\nu_{e}}{2}\right)^2},
\end{equation}
\begin{equation}
\label{Minority_Branch}
\omega_{-} =-i\left(\frac{\nu_h}{2}+\frac{\nu_{eh}}{3}\right)+\sqrt{{{k}^{2}}v_h^{2}(1+2r_h)^2- \left(\frac{\nu_h}{2}+\frac{\nu_{eh}}{3}\right)^2}.
\end{equation}

A detailed analysis of these solutions shows that the branches $\omega_+$ and $\omega_-$ correspond to the oscillations of majority (electrons) and minority (holes) carriers, respectively. The minority carrier oscillations are strongly damped by electron-hole friction and, therefore, they are omitted in the further consideration.

The wave velocity of the electronic plasma oscillations $s_+$ is  
\begin{equation}
\label{Plasma_velocity}
s_+=v_e(1+r_e)^{1/2}\simeq v_F\sqrt{4\alpha k_F d} \propto V_G^{1/4}d^{1/4},
\end{equation}
where $\alpha=e^2/\kappa\hbar v_F$ is the coupling constant ("fine-structure constant") and $k_F=\mu_e/\hbar v_F$ is the Fermi momentum of electrons. This velocity markedly exceeds the Fermi one and is primarily determined by the Coulomb interaction of carriers with the gate.

The dispersions for electron-hole sound in intrinsic graphene and plasma waves in gated graphene at high gate voltages are depicted in Fig. 2 in the THz range of frequencies. The curves are calculated under the assumption of equal chemical potentials $\mu_{e,0}=\mu_{h,0}$, i.e. when the non-perturbed state is an equilibrium one. The spectra exhibit non-linear dispersion at low frequencies originating from the damping (see the inset in Fig. 2).

\begin{figure}
\includegraphics[width=1\linewidth]{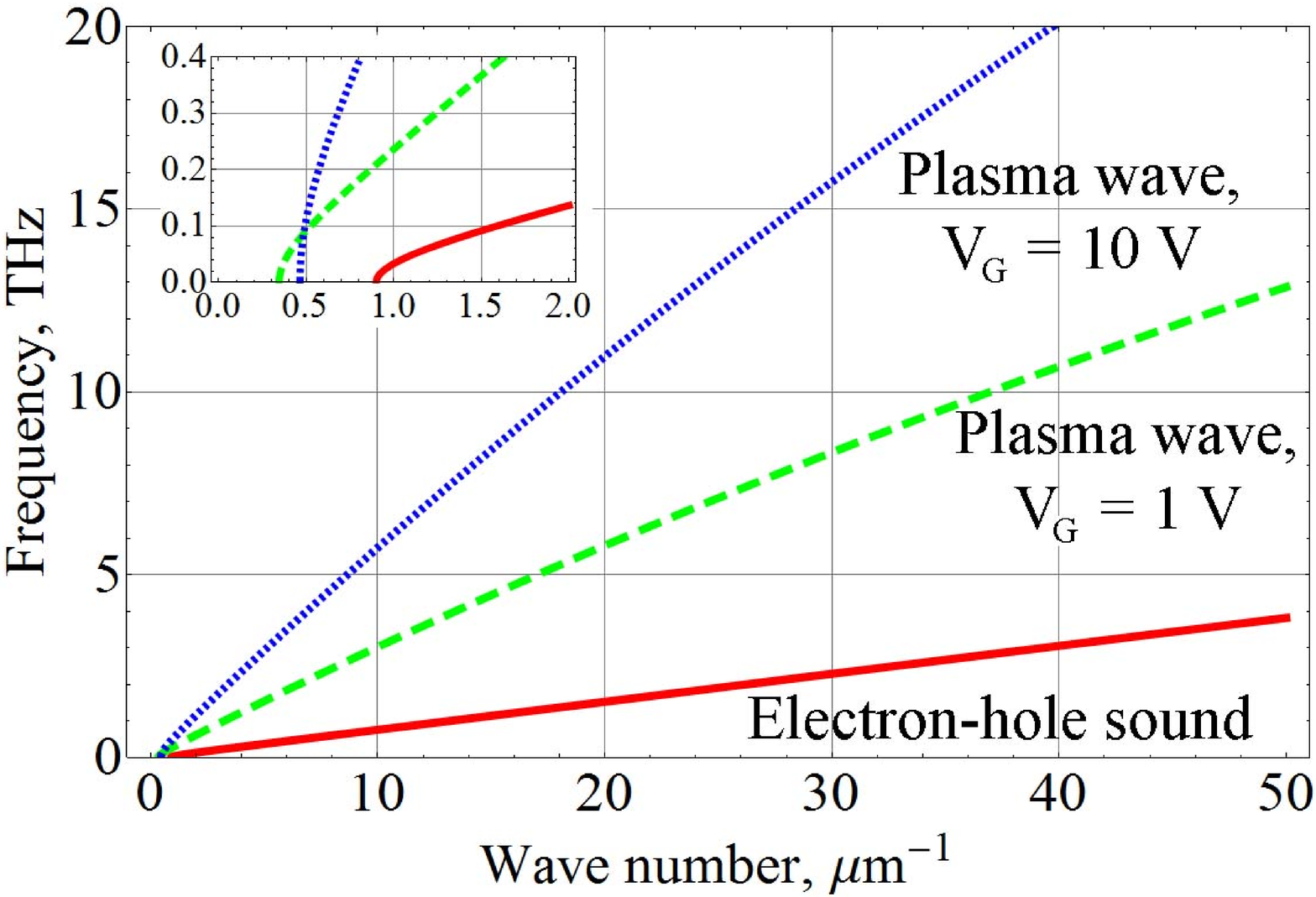}
\caption{\label{Fig2}Dispersion of plasma and electron-hole sound waves at different gate voltages. Inset: non-linear dispersion of the waves at small frequencies resulting from collisions}
\end{figure}

Worth mentioning the dispersions are also non-linear at short wavelengths ($kd \sim 1$) when the gradual channel approximation becomes inapplicable. However, this limitation can be overcome if we use the rigorous solution of Poisson equation~\cite{Ryzhii-plasma-2} instead of Eq.~(\ref{Gradual_ch}). For the considered gated structure the transition to the rigorous solution can be performed by the substitution:
\begin{equation}
\label{Rigorous_Poisson}
C\rightarrow \frac{2\kappa k}{4\pi (1-e^{-2kd})}.
\end{equation}

Equation~(\ref{Rigorous_Poisson}) restrains the unlimited growth of the plasma wave velocities for large distances $d$ between graphene sheet and metal gate.

\subsection{Velocities and damping rates of the waves}
The wave velocities and damping rates can be calculated analytically in the linear domain of spectra, i.e. at ($\nu_{e,h} < \omega < s/d $) at an arbitrary value of electron and hole densities. Assuming $\omega = s k$, where $s$ is the wave velocity, and plugging this into Eq.~(\ref{General_dispersion}), one obtains two solutions $s_-$ and $s_+$:
\begin{multline}
\label{General_velocity}
s_{\pm}^2= \frac{1}{2}  [v_e^2(1+r_e)+v_h^2(1+r_h)] \pm \\
\frac{1}{2}\sqrt{[v_e^2(1+r_e)-v_h^2(1+r_h)]^2+4v_e^2v_h^2r_er_h},
\end{multline}
consistent with Eqs.~(\ref{Sound_velocity},~\ref{Plasma_velocity}). 

\begin{figure}
\includegraphics[width=1\linewidth]{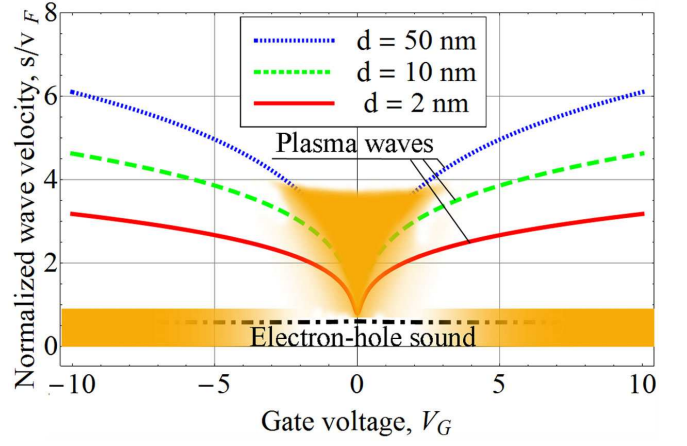}
\caption{\label{Fig3}Velocities of plasma waves vs. gate voltage calculated for different gate layer thicknesses [Eq.~\ref{Plasma_velocity}]. Dash-dotted line corresponds to the electron-hole sound velocity in the vicinity of the neutrality point. Regions of strong damping are filled}
\end{figure}

In the vicinity of Dirac point the waves with the lower velocity $s_-$ correspond to the electron-hole sound, while in the monopolar plasma they turn into the oscillations of minority carriers. The waves with the higher velocity $s_+$ behave as plasma waves at any electron and hole densities. The solutions given by Eq.~(\ref{General_velocity}) are plotted in Fig.~3 as the functions of gate voltage with the assumption of equal steady-state chemical potentials ($\mu_{e,0}=\mu_{h,0}$). One can readily see that the velocity $s_-$ almost does not depend on the gate voltage applied, while the velocity $s_+$ exhibits unlimited growth at high gate voltages $s_+ \propto V_G^{1/4}$. 

\begin{figure}
\includegraphics[width=1\linewidth]{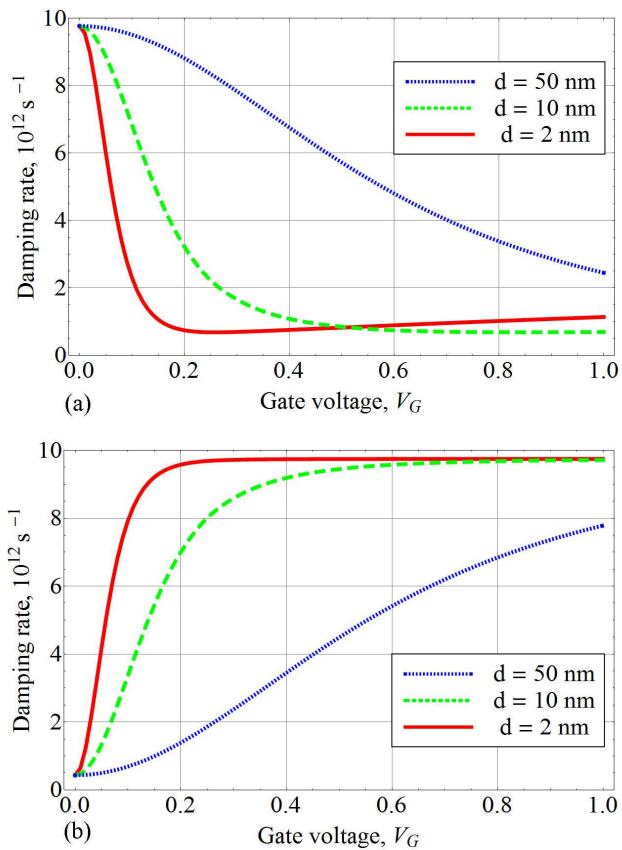}
\caption{\label{Fig4}Damping rates for the two branches of spectrum: $\omega_+$ (a, top) and $\omega_-$ (b, bottom)}
\end{figure}

To calculate the damping rates of the waves we assume $\omega_\pm=s_{\pm}k+i\gamma_{\pm}$, where $\gamma_{\pm}$ characterizes the damping rate of the waves. The obtained damping rates $\gamma_\pm$ vs. the gate voltage are demonstrated in Fig.~\ref{Fig4}. It is clearly seen that in symmetric bipolar plasma the electron-hole sound branch $\omega_-$ exhibits weak damping. In the monopolar case ($\Sigma_e\gg\Sigma_h$) the plasma wave, corresponding to the oscillations of the majority carriers, is weakly damped. The minimum damping rate for the waves considered is of the order of $5 \times 10^{11}$~s$^{-1}$ and is determined by electron-phonon (or hole-phonon) collision frequency if the only scattering mechanism is acoustic phonon scattering. At high gate voltages the damping coefficient $\gamma_+$ grows linearly in accordance with the expression for characteristic collision frequency $\nu_{e}$. The maximum damping rate of the waves is of the order of electron-hole collision frequency $\nu_{eh}$. In accordance with these calculations, the regions of strong damping in Fig.~\ref{Fig3} are marked with filling.

\subsection{Comparison with other models}
Plasma waves in gated graphene were discussed in Refs.~\cite{Ryzhii-plasma-1, Ryzhii-plasma-2} using the kinetic approach under assumption of collisionless transport. As expected, formally setting the collision frequency zero in Eq.~(\ref{Plasma_branch}), we obtain the spectrum of plasma waves, derived previously~\cite{Ryzhii-plasma-2}. Meanwhile, in the kinetic approach the existence of the electron-hole sound could not be predicted as it was assumed that the wave velocity should overcome Fermi velocity.

\begin{figure}
\includegraphics[width=1\linewidth]{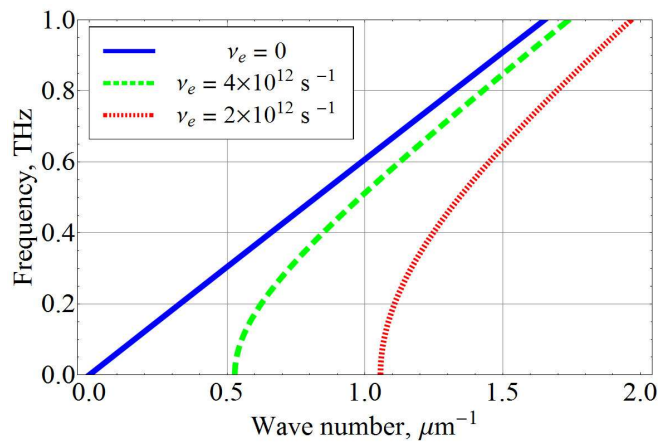}
\caption{\label{Fig5}Comparison of dispersions for plasma waves, calculated using 
kinetic model~\cite{Ryzhii-plasma-2} (disregarding collisions with impurities and phonons, solid line)
and hydrodynamic model for $V_G=10$~V, $d=10$~nm, $\kappa=4$, and three different collision frequencies.}
\end{figure}

The dispersion curves of the plasma waves obtained via the hydrodynamic and kinetic theories are compared in Fig. 5 for the same parameters of the gated structure ($\kappa=4$, $d=10$ nm), gate voltage $V_G=10$ V and three different collision frequencies. The only difference between the curves is in the  non-linear part of spectra originating from collisions.

Plasmons in graphene were also intensively studied within the analysis of the polarizability function $\Pi(q,\omega)$  in random-phase approximation \cite{Plasmons-RPA-1, Plasmons-RPA-2, Soundarons}. In particular, the square-root $\omega\propto\sqrt{k}$ and linear $\omega\propto k$ dispersions were obtained for the non-gated and gated graphene, consequently. To obtain the plasma wave dispersion for the non-gated graphene in hydrodynamic model we tend $d$ to infinity in Eq.~(\ref{Rigorous_Poisson}) and plug it into Eq.~(\ref{Plasma_branch}). Thus, plasma waves in non-gated graphene at high electron densities exhibit the following dispersion:
\begin{equation}
\label{Dispersion_no_gate}
\omega_{+} = \sqrt{\frac{v_F^2k^2}{2} \left( 1 + 4\alpha k_F \right)-\left(\frac{\nu_{e}}{2}\right)^2} - i \frac{\nu_{e}}{2}.
\end{equation}
In the low frequency limit the above equation is simplified
\begin{equation}
\label{RPA_limit}
\omega_{+} \simeq v_F\sqrt{2\alpha k k_F},
\end{equation}
and the result quantitatively coincides with that predicted in~\cite{Plasmons-RPA-2}. What concerns the result obtained in \cite{Soundarons} for plasmons in gated graphene, it coincides both with that obtained within kinetic approach \cite{Ryzhii-plasma-2} and, hence, with our result~(\ref{Plasma_branch}).

\begin{figure}
\includegraphics[width=1\linewidth]{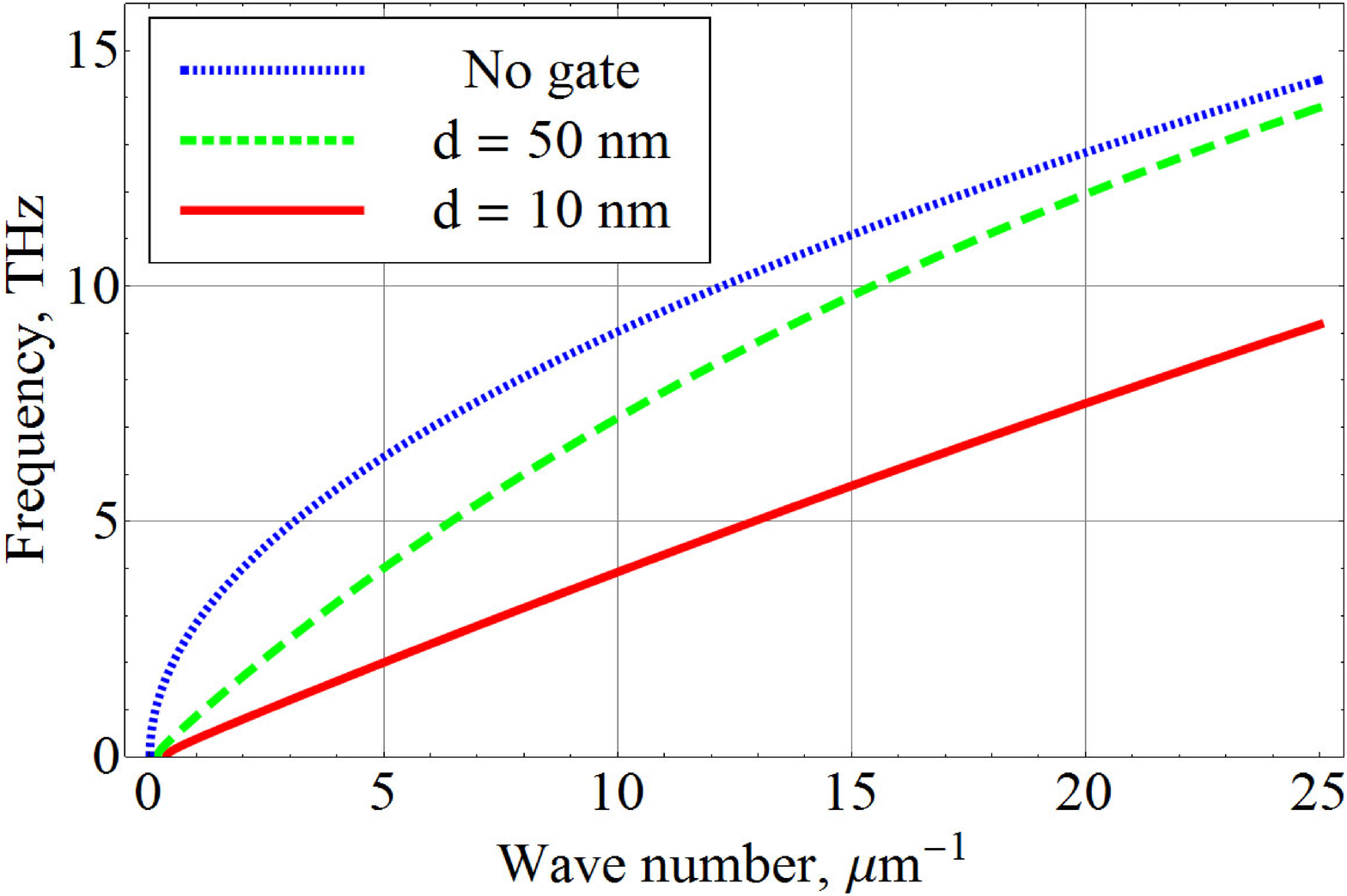}
\caption{\label{Fig6} Dispersions of plasma waves at a given carrier density $\Sigma_e = 5\times 10^{12}$ cm$^{-2}$ and different thicknesses of gate dielectric}
\end{figure}

The transition from almost linear to square-root plasma wave dispersion at different gate dielectric thicknesses is shown in Fig.~\ref{Fig6} for the given value of electron density $\Sigma_e = 5\times 10^{12}$ cm$^{-2}$. In the short-wave limit $kd\gg 1$ (providing the applicability of the hydrodynamic model) the dispersion of plasma waves is linear again with the velocity depending only on the fundamental constants of graphene.  Worth mentioning that the electron-hole sound waves are insensitive to the gate and exhibit almost linear dispersion at any gated structure parameters.

\section{Discussion of the results}
The hydrodynamic model of electron and hole transport in graphene which takes into account the linearity of carrier spectra has been developed. Strong interactions between electrons and holes result in mutual frictional forces proportional to the mismatch in electron and hole drift velocities. The interaction of carriers with acoustic phonons and impurities is also governed by a friction term in Euler equations. The estimated electron-hole collision rate is greater than that between carriers and phonons (impurities) at room temperature.
 
The model has been applied to the solution of two challenging problems: dc graphene conductivity and spectra and damping of the collective excitations in graphene structures.

The hydrodynamic equations derived provide an opportunity to calculate the conductivity of graphene sheet in a wide range of gate voltages including the Dirac point, where the influence of electron-hole collisions on the charge transport is dominant. However, when there is even a small difference (proportional to the ratio of external scattering rate to that of carrier-carrier) between electron and hole densities the majority carriers drag the minority ones. In this case, the conductivity is governed by scattering on phonons and impurities. The effect of drag is particularly pronounced in high-purity (suspended) graphene samples.

The spectra of collective excitations in electron-hole system in graphene have been calculated. The existence of two types of the excitations has been revealed: quasineutral electron-hole sound waves and plasma waves. 

In the vicinity of the Dirac point or in the optically pumped intrinsic graphene the sound waves undergo weak damping of the order of  $5\times 10^{11}$~s$^{-1}$ in perfect structures. The damping of such waves is caused by weak scattering on acoustic phonons and is insensitive to strong electron-hole scattering. The latter fact is due to co-directional motion of electrons and holes in quasi-neutral sound waves, which almost eliminates electron-hole friction.

The quasi-neutral electron-hole sound waves under consideration are akin to those in bipolar electron-hole plasma in semimetals considered by Konstantinov and Perel a long time ago~\cite{Konstantinov-Perel}. Later such waves were studied in semiconductors~\cite{35}, two-band metals~\cite{36} and superconducting compounds~\cite{37}. Recently the acoustic plasma waves were studied in two dimensional semimetals~\cite{Chaplik} (CdHgTe/HgTe/CdHgTe quantum wells).  It was shown that those waves can contribute to the thermodynamic properties of the materials, e.g. carrier relaxation rates in bulk semiconductors~\cite{38} and critical temperatures of superconducting compounds~\cite{37}. In graphene such waves could be feasibly excited by optical spots or modulated optical radiation. In both situations quasi-neutral non-uniform distributions of electrons and holes arise.
 
The propagation of plasma waves in symmetric electron-hole systems is strongly suppressed by electron-hole collisions. Strong damping of such waves can significantly decrease the rate of electron-hole recombination via plasmon emission, discussed in Ref.~\cite{Rana-recomb} In its turn, the recombination via the emission of electron-hole sound waves is prohibited as their velocity is smaller than $v_F$ [Eq.~(\ref{General_velocity})].

In sufficiently monopolar systems (electron or hole plasma, e.g. in gated graphene at high gate voltages) the damping of plasma waves, corresponding to the oscillations of majority carriers, is associated solely with the scattering on disorder (phonons and impurities). It can be rather weak in perfect structures, particularly at low temperatures. The dispersion of plasma waves in gated graphene structures is linear in a wide range of frequencies~\cite{Chaplic-Old}. The dependence of the plasma wave velocity on the gated structure parameters is quantitatively similar to that obtained for the two-dimensional electron gas of massive electrons~\cite{Dyakonov-Shur}.

The hydrodynamic and kinetic models yield the same dispersions of plasma waves if one formally tends the collision frequency to zero. However, the hydrodynamic approach provides a regular way to describe the damping of the waves, associated with carrier scattering.

\textit{In conclusion}, the hydrodynamic equations for bipolar graphene system were derived and applied for calculation of dc conductivity and spectra of collective excitations in graphene structures. The model opens the prospects to the simulation of graphene-based transistors, THz-range detectors and generators, and light emitting devices.

\section*{Acknowledgment}
The research was supported via the grants 11-07-00464-a of the Russian Foundation for Basic Research, F793/8-05 of Computer Company NIX (science@nix.ru), and by the Japan Science and Technology Agency, CREST, Japan.

\section*{Appendix 1. Dissipative terms in the Euler equations}
\setcounter{equation}{0}
\renewcommand{\theequation} {A\arabic{equation}}

In this appendix we derive explicit expression for the friction forces, associated with carrier-carrier, carrier-impurity, and carrier-phonon scattering. 

\subsection{Mutual electron-hole friction}
To derive the expression for electron-hole friction force one should linearize the collision integral
\begin{multline}
St\{f_e,f_h\}=\\*
4 \int W(q)\{f_e({\bf p-q}){f_h}({\bf p}_{1}+{\bf q})
[ 1- f_e({\bf p })][ 1-f_h({\bf p}_1)] -\\*
\label{E_H-coll-int} -{f_e}({\bf p}) {f_h} ( {\bf p}_1) [ 1-{f_e} ({\bf p}-{\bf q})] [ 1-{f_h} ({\bf p}_1+{\bf q} ) ] \}  \frac{d^2{\bf p}_1 d^2{\bf q}}{(2\pi\hbar)^4},
\end{multline}
where the non-equilibrium distribution functions for electrons and holes are determined via Eqs.  (\ref{F_el}) and (\ref{F_hole}), the factor of 4 arises from 2 possible spins and valleys for particle ${\bf p_1}$. $W(q)$ is the Coulomb scattering probability which can be written down as
\begin{multline}
W(q) = \frac{2\pi}{\hbar v_F}\left[ \frac{2\pi \hbar e^2}{\kappa (q+q_{TF})}\right]^2 \left|\langle u_{\bf p}|u_{\bf p'}\rangle\right|^2 \left|\langle u_{{\bf p}_1}|u_{{\bf p'}_1}\rangle\right|^2 \times \\
\label{Coul_probability} \times \delta\left(p+p_1-|{\bf p} -{\bf q}|+|{\bf p}_1+{\bf q}|\right) 
\end{multline}
where $q_{TF}=4 \alpha T/v_F\ln(1+e^{\mu_e/T})(1+e^{-\mu_h/T})$ is the Thomas-Fermi momentum describing screening in graphene \cite{Electronic-properties, Plasmons-RPA-2}, $\left|\langle u_{\bf p}|u_{\bf p'}\rangle\right|^2=(1+\cos\theta_{pp'})/2$ is the matrix element of two Bloch functions in honeycomb lattice. In Eq.~(\ref{Coul_probability}) the Thomas-Fermi momentum could be replaced by the reciprocal gate-dielectric thickness for the qualitative description of gate screening.

In case of small velocities: ${\bf p}\cdot {\bf V}_e\ll T$, ${\bf p}\cdot {\bf V}_h\ll T$ the electron-hole collision integral can be transformed using the common linearization technique for Fermi-systems \cite{Kinetics}:
\begin{multline}
St\{f_e,f_h\} =\\
4 \int W(q) f_e({\bf p}) f_h ({\bf p}_1) [ 1-f_e({\bf p}-{\bf q} )] [1- f_h( {\bf p}_1 +\bf q)]\times \\
\label{Stoss} \frac{ {\bf q}\cdot ({\bf V}_h-{\bf V}_e)}{T} \frac{d^2{\bf q} d^2{{\bf p}_1}}{(2\pi\hbar)^4},
\end{multline}
where the energy and momentum conservation laws were used.

Eq.~(\ref{Stoss}), timed by ${\bf q}/2$ and integrated over $d\Gamma_{\bf p}$ is electron-hole friction force ${\bf f}_{eh}$ per unit area in the right hand side of the Euler equations~(\ref{Euler_el}, \ref{Euler_hole}). After averaging over angle between $\bf q$ and ${\bf V}_h-{\bf V}_e$ it is presented as
$$
{\bf f}_{eh}=\beta_{eh} \left( {\bf V}_h-{\bf V}_e \right),
$$
where
\begin{multline}
\beta_{eh}= \int W (q) {f_e} ( {\bf p} ) {f_h} ({\bf p}_1)  [ 1-{f_e} ({\bf p}-{\bf q} )] [ 1-{f_h} ( {\bf p}_1 + {\bf q})] \times\\
\frac{4 q^2}{T} \frac{ d^2{\bf q} d^2{\bf p}_1 d^2{\bf p}}{(2 \pi \hbar)^6}.
\end{multline}

In~\cite{Quantum-critical, Kashuba} it was shown that almost collinear vectors ${\bf p}$, ${\bf p_1}$ and ${\bf q}$ give the leading contribution to the collision integral. This facts originates from the linearity of the carrier spectrum, i.e. particles moving with the same velocity and the same direction interact infinitely long. Following the technique, described in~\cite{Kashuba}, the quantity $\beta_{eh}$ in the collinear limit is presented as
\begin{widetext}
\begin{multline}
\label{Collinear_limit} \beta_{eh} \propto \frac{T^4 e^4 \ln(1/\alpha)}{\hbar^5 \kappa^2 v_F^6}  \int_0^{\infty}dP\int_0^{\infty}dP_1\int_{-P_1}^{P}
F(P-z_e) F(P_1+z_h) \left[ 1-F(P - Q - z_e)\right]  \times \\
\left[ 1-F(P_1 + Q+ z_h)\right] \sqrt{PP_1( P - Q)(P_1 + Q)} \frac{ Q^2 dQ}{(|Q| - 4\alpha  \ln F(z_e)F(-z_h))^2},
\end{multline}
\end{widetext}
where $F(x)=\left(1+e^x\right)^{-1}$, $z_e=\mu_e/T$, $z_h=\mu_h/T$, and the term $\ln 1/\alpha$ originates from the corrections to the electronic sperctrum due to electron-electron interactions. In this form the friction coefficient $\beta_{eh}$ can be simply computed numerically.

\subsection{Friction caused by charged impurities and phonons}

Calculation of friction force caused by charged impurities can be drawn analytically in the limit of monopolar plasma. The momentum relaxation rate is given by:
\begin{equation}
\label{EqA4}
\tau_{p,i}^{-1}=\int \Sigma_i W(q) (1-\cos\theta_{pp'}) \left|\langle u_{\bf p}|u_{\bf p'}\rangle\right|^2 \frac {d^2{\bf{p'}}}{(2\pi\hbar)^2},
\end{equation}
where $\Sigma_i$ is the sheet density of charged impurities. The Coulomb scattering probability $W(q)$ is given by eq.~(\ref{Coul_probability}) with the following delta-function: $\delta(pv_F-p'v_F)$. Here we put into consideration only self-screening caused by carriers in graphene.

To calculate the friction force per unit area ${\bf f}_{ei}$ one should integrate the momentum relaxation rate~(\ref{EqA4}) timed by ${\bf p}$ with the non-equilibrium part of distribution function
$\delta f_e=- \frac{\partial f_{e,0}}{\partial \varepsilon} {\bf p} \cdot {\bf V}_e$:
\begin{equation}
\label{Force_i}
{\bf{f}}_{ei}=\int {\bf p}\frac{\partial f}{\partial \varepsilon} {\bf p} \cdot {\bf V}_e \tau_{p,i}^{-1} d\Gamma_{\bf{p}}.
\end{equation}
In monopolar case one can set $\partial f_{e,0}/\partial\varepsilon=-\delta(pv_F-\mu_e)$ and perform trivial integration:
$$
{\bf f}_{ei}=-\frac{\pi e^4 \Sigma_e\Sigma_i {\bf V}_e}{\hbar v_F^2\kappa^2}\int_0^{2\pi} \frac{\sin^2\theta_{pp'}d\theta_{pp'}}{(2\sin(\theta_{pp'}/2)+4\alpha)^2}
$$
where the dimensionless integral actually depends only on the permittivity $\kappa$ of the environment.

We can use the same technique to derive the expression for the friction force, caused by phonon scattering. We start from the expression
$$
\tau_{p,ph}^{-1}=\frac{D^2 T p}{4\rho_ss^2\hbar^3v_F},
$$
derived in~\cite{Vasko-Ryzhii}, and, using Eq.~(\ref{Force_i}), arrive at the following friction term:
\begin{equation}
{\bf f}_{e\,ph}=-\frac{D^2 T\langle p_e^2\rangle {\bf V}_e}{4\rho_s s^2 v_F^2 \hbar^3}.
\end{equation}

\bibliography{biblio/svintsov_plasma_10}

\end{document}